# Magneto-conductance Oscillations in Graphene Antidot Arrays

T. Shen,[1,2] Y. Q. Wu, [1] M.A. Capano, [1] L.P. Rokhinson,[2] L.W. Engel,[3] and P.D. Ye [1,a)]
[1)] *School of Electrical and Computer Engineering and Birck Nanotechnology Center, Purdue University, West Lafayette, IN 47907*
[2)] *Department of Physics, Purdue University, West Lafayette, IN 47907*
[3)] *National High Magnetic Field Laboratory, Tallahassee, FL 32310*

Epitaxial graphene films have been formed on the C-face of semi-insulating 4H-SiC substrates by a high temperature sublimation process. Nano-scale square antidot arrays have been fabricated on these graphene films. At low temperatures, magneto-conductance in these films exhibits pronounced Aharonov-Bohm oscillations with the period corresponding to magnetic flux quanta added to the area of a single antidot. At low fields, weak localization is observed and its visibility is enhanced by intravalley scattering on antidot edges. At high fields, we observe two distinctive minima in magnetoconductance which can be attributed to commensurability oscillations between classical cyclotron orbits and antidot array. All mesoscopic features, surviving up to 70 K, reveal the unique electronic properties of graphene.

[a)]Author to whom correspondence should be addressed; electronic mail: yep@purdue.edu

Antidot arrays are interesting structures to study because of their transport properties in conducting electronic materials. [1] Antidots can be regarded as groups of the imposed scatters that limit the ballistic transport or the mean free path of carriers. Antidot arrays can also be considered as an ensemble of Aharonov-Bohm (AB) rings connected together. [2-3] Moreover, antidot arrays are ideal for investigating quantum coherent effects and phase coherence length of carriers. Graphene, a monolayer of carbon atoms tightly packed into a two-dimensional (2D) hexagonal lattice, has recently been shown to be thermodynamically stable and exhibits astonishing transport properties, such as an electron mobility of ~15,000 $cm^2/Vs$ and electron velocity of ~$10^8$ cm/s at room temperature. [4-9] However, graphene is semi-metallic and thus not suitable for most electronic and optoelectronic applications which require a semiconductor with a specific, finite bandgap. Antidot arrays impose lateral potential barriers that could create a bandgap in graphene, [10] similar to the creation of the energy gap by introducing lateral periodic potentials from positive ion cores in a real semiconductor crystal. In this paper, the magneto-transport properties of this kind of nano-structured antidot arrays in epitaxially grown graphene film on SiC are investigated. Pronounced AB oscillations, weak localization, and commensurability oscillations are observed, directly related to ballistic transport properties, such as mean free path, and coherent transport properties, such as phase coherence length.

The advantage of epitaxially grown graphene for nanoelectronic applications resides in its planar 2D structure that enables conventional top-down lithography and processing techniques. [7-9] The graphene films in this Letter are grown on the carbon face of semi-insulating 4H-SiC substrates in an Epigress VP508 SiC hot-wall chemical vapor deposition (CVD) reactor. The off-cut angle of the substrate is nominally zero degrees. Prior to growth, substrates are subjected to a hydrogen etch at 1600 $^{o}$C for 5 minutes, followed by cooling the samples to below 700 $^{o}$C. After evacuating hydrogen from the system, the growth environment is pumped to an approximate pressure of $2\times10^{-7}$ mbar before temperature ramping at a rate of 10-20 $^{o}$C/min and up to a specified growth temperature. Growth temperature and time for this particular graphene film are 1550 $^{o}$C for 10 minutes. A room-temperature field-effect mobility measured from graphene grown under these conditions is as high as 5400 $cm^2$/Vs .[8]

The device structure of the fabricated graphene antidote array is shown in Fig. 1(a) and (b). Device isolation and antidot formation of the graphene film is realized by $O_2$ plasma based dry etching with electron-beam-lithographically (Vistec VB-6 UHR-EWF) defined HSQ resist as the protection layer. The diameter of the holes is around 40 nm and the defined antidot array period is 80 nm. Ti/Au metallization is used to form the two terminal Ohmic contacts on graphene film. Two-point resistance measurements are performed in a variable temperature (0.4K to 70K) $^{3}$He cryostat in magnetic fields up to 18 T using low frequency lock-in technique. The external magnetic field ($B$) is applied normally to the graphene plane.

Figure 1(c) shows the magneto-conductance $G(B)$ of graphene antidot arrays as a function of perpendicular magnetic field at 0.47 K. The trace is essentially symmetric, $G(B)=G(-B)$, which is the reciprocity relation mandatory for a two–terminal measurement of a stable device. There are three distinguishing features of the measured magneto-conductance. The first is the pronounced weak localization dip around zero magnetic field. The second are distinct conductance minima at ±4T and ±8T. We attribute these minima to the commensurability between the cyclotron orbits of carriers in certain magnetic fields and the period of artificial holes as illustrated in Figure 1(b). The magneto-conductance shows minima when the trajectories of carriers are trapped in this antidot array. Using the commensurability relation $2R_c=2\sqrt{(\pi N_s)}(h/2\pi eB)=a$, carrier density $N_s$ is determined to be ~ $7.5\times10^{12}/cm^2$, where $R_c$ is the cyclotron radius, $h$ is the Planck constant, and $a$ is the period of antidot arrays. The extraordinary long mean free path or the ballistic trajectory of cyclotron orbits is not fully understood. It could be related with the absence of back scattering mechanism in graphene films. [11] The third feature is the tiny structures superimposed on the measured trace, for example, between ±2T - ±7T. Universal conductance fluctuations are suppressed due to assembling of thousands of holes with different phases at each antidote. These tiny periodic features are identified as AB oscillations, which are related with each magnetic quantum flux penetrating in one antidot cell. AB oscillations on an exfoliated graphene film have been demonstrated experimentally on a single lithographically defined ring. [12] Universal conductance fluctuations in our sample are strongly suppressed due to assemble averaging over the large array.

The black thick curve in Figure 2(a) shows the measured magneto-conductance between +2 T – +12 T with superimposed oscillatory features. By taking the difference of the measured curve and the black baseline curve obtained from smoothing the measured one, a pronounced AB periodically oscillatory curve is exhibited as shown as the grey curve in Figure 2(a). The AB oscillation period $\Delta B \approx 0.5$ T in this field range is consistent with the condition that the magnetic flux enclosed within the unit cell of the square antidot lattice changes by a single magnetic flux quantum, i.e., $\Delta B = (h/e)/a^2$ with $a$=80 nm. The rms amplitude of AB oscillations is ~ 0.01 $e^2/h$. For detailed discussions, Figure 2(b) illustrates Fourier power spectrum of Figure 2(a) with a broad peak centered around 0.47 T, with 0.57 T and 0.40 T as the edge of the half-height width. It corresponds to the inner radius, middle radius, and outer radius of 48 nm, 53 nm, and 57 nm, respectively, if $\Delta B = (h/e)/(\pi r^2)$ where $\pi r^2$ is associated with the effective antidot area. It is consistent with the designed geometry well with 40 nm holes and 80 nm pitches. The relatively large inner radius could be related with overdeveloped resist patterns, plasma over-etching and certain depletion length of graphene edges with unpassivated dangling bonds. The magnetic length [$l_B=\sqrt{h/(2\pi eB)}$=9.2 nm at $B$=6T] or similar edge channels in the quantum Hall regime could also affect the data. The observed AB oscillations demonstrate that the epitaxial graphene on SiC is of high-quality and at least has the quantum coherent length larger than 80-100 nm. The weak peak features around 1/B = 4 (1/T) could be related to *h/2e* oscillations. [3,12]

While universal conductance fluctuations are generally observed in small graphene flakes, weak localization correction is strongly reduced compared to the conventional two-dimensional (2D) systems due to inherently suppressed backscattering in graphene [11,13]. Scattering on impurities or sharp edges introduces intravalley scattering [14], which restores weak localization corrections [15]. We observe pronounced negative magneto-resistance at low fields with a sharp cusp at zero field characteristic of weak localization in 2D, see Figure 3(a). Moreover, at higher fields magneto-resistance changes sign, which is expected for the case of strong intravalley scattering [16]. We used theory developed in [16] to analyze the data and extract both phase coherence length $L_\phi$ and intravalley scattering length $L_i$, Figure 3(b). $L_i$ is found to be temperature independent and is approximately equal to the distance between antidots, suggesting that scattering on the antidot edges is the dominant intravalley scattering mechanism in our samples. $L_\phi$ decreases with the increasing temperature, although it does not follow 1/T dependence found in unpatterned graphene [14]. We also note that the range of field where weak localization is observed in antidot array is much larger than that for the unpatterned samples. The temperature dependence of AB oscillations is also plotted in Figure 3(c), which is consistent with the conclusion from Figure 3(b) by weak localization peaks fitting.

In conclusion, we present magneto-transport experiments on antidot arrays fabricated on epitaxially grown graphene films on SiC. The experiment demonstrates the observation of commensurability oscillations and AB oscillations arising from the artificially imposed lateral potential modulation. It opens up a new possibility to engineer bandgap in graphene films for future device applications.

The authors would like to thank J.A. Cooper Jr., Y. Lyanda-Geller, M.S. Lundstrom, T. Low, J. Appenzeller, Kun Xu for valuable discussions, and G. Jones, T. Murphy and E. Palm at National High Magnetic Field Laboratory (NHMFL) for experimental assistance. Part of the work on graphene is supported by NRI (Nanoelectronics Research Initiative) through MIND (Midwest Institute of Nanoelectronics Discovery), The Indiana 21st Century Fund, DARPA and Intel Cooperation. NHMFL is supported by NSF Grant Nos. DMR-0084173 and ECS-0348289, the State of Florida, and DOE.

**Figure Captions**

Figure 1 (a) The sample layout with 2μm × 2μm graphene area and two-terminal metal contacts. (b) Electron microscopic image of antidote arrays with ~40 nm holes and ~80 nm pitches. The commensurate orbits around 1 antidot and 4 antidots are sketched to illustrate the physical origin of Weiss oscillations. (c) Magnetoconductance of the graphene antidote arrays measured at T=477 mK. On top of the Weiss oscillations, periodic features are clearly visible as also highlighted in Figure 2(a).

Figure 2 (a) The solid curve is the measured magnetoconductance. The thin curve is the “baseline” after smoothing the original measured curve. The periodically oscillatory curve is the subtraction of the two black curves. Vertical straight lines are guided by eyes showing periodic *B* feature of observed AB oscillations. (b) Fourier spectrum of the oscillatory grey curve between 2T to 12 T. The solid curve is after 6 points smoothing. The two vertical straight lines with arrows indicate the positions for half-height of the observed *h/e* peak, as used for the calculation of the inner and outer radii of AB-“ring” structure around one antidot.

Figure 3 (a) Magneto-resistance as a function of the sample temperature from 477 mK up to 70 K. The solid black curves are fitted weak localization curves in antidot arrays [16]. (b) The coherence length $L_{\Phi}$ and intravalley scattering length $L_{\mathrm{i}}$ of graphene film with antidot arrays versus sample temperatures. (c) Temperature dependence of AB oscillations. The traces are vertically shifted for clarity.

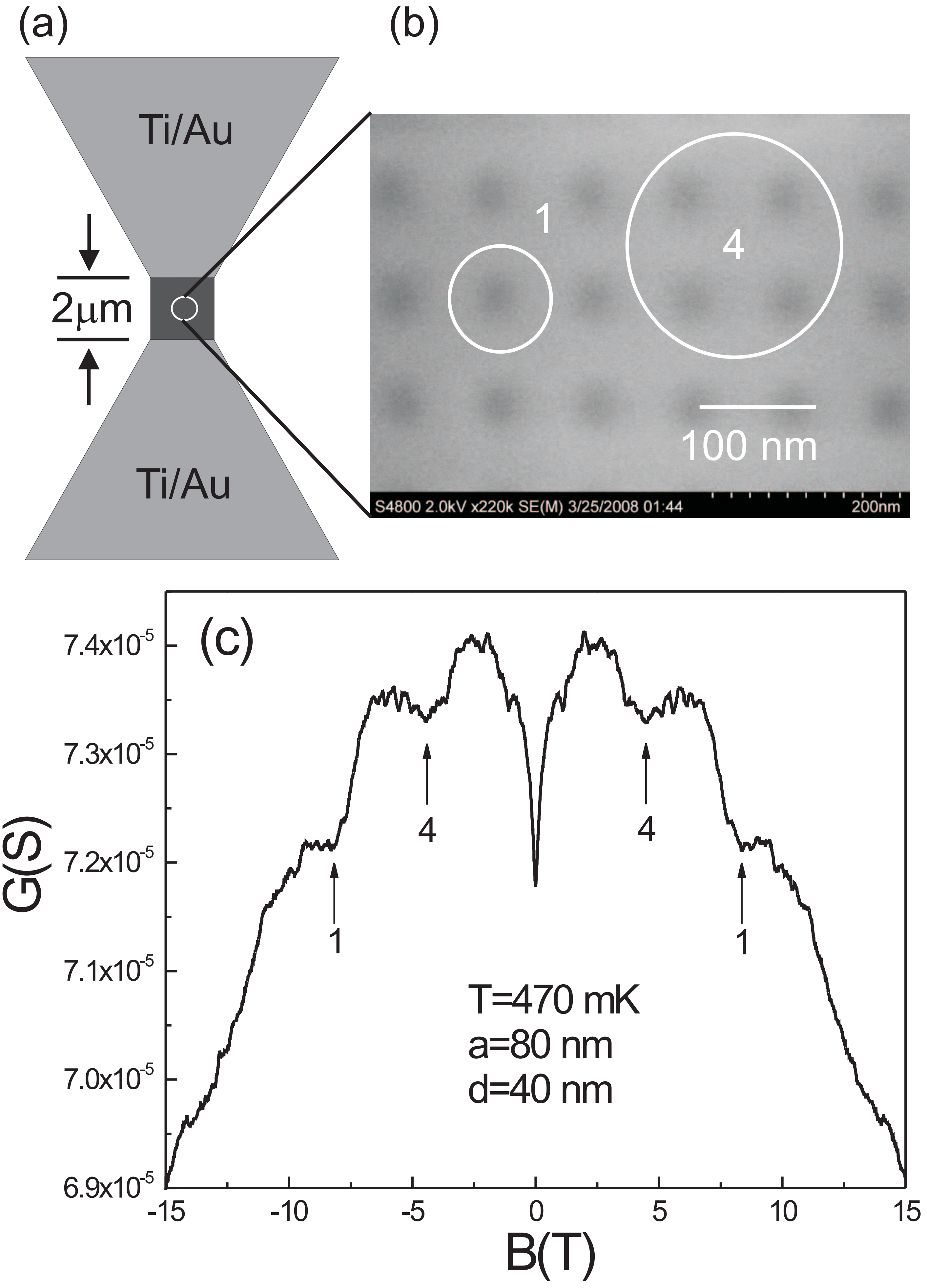

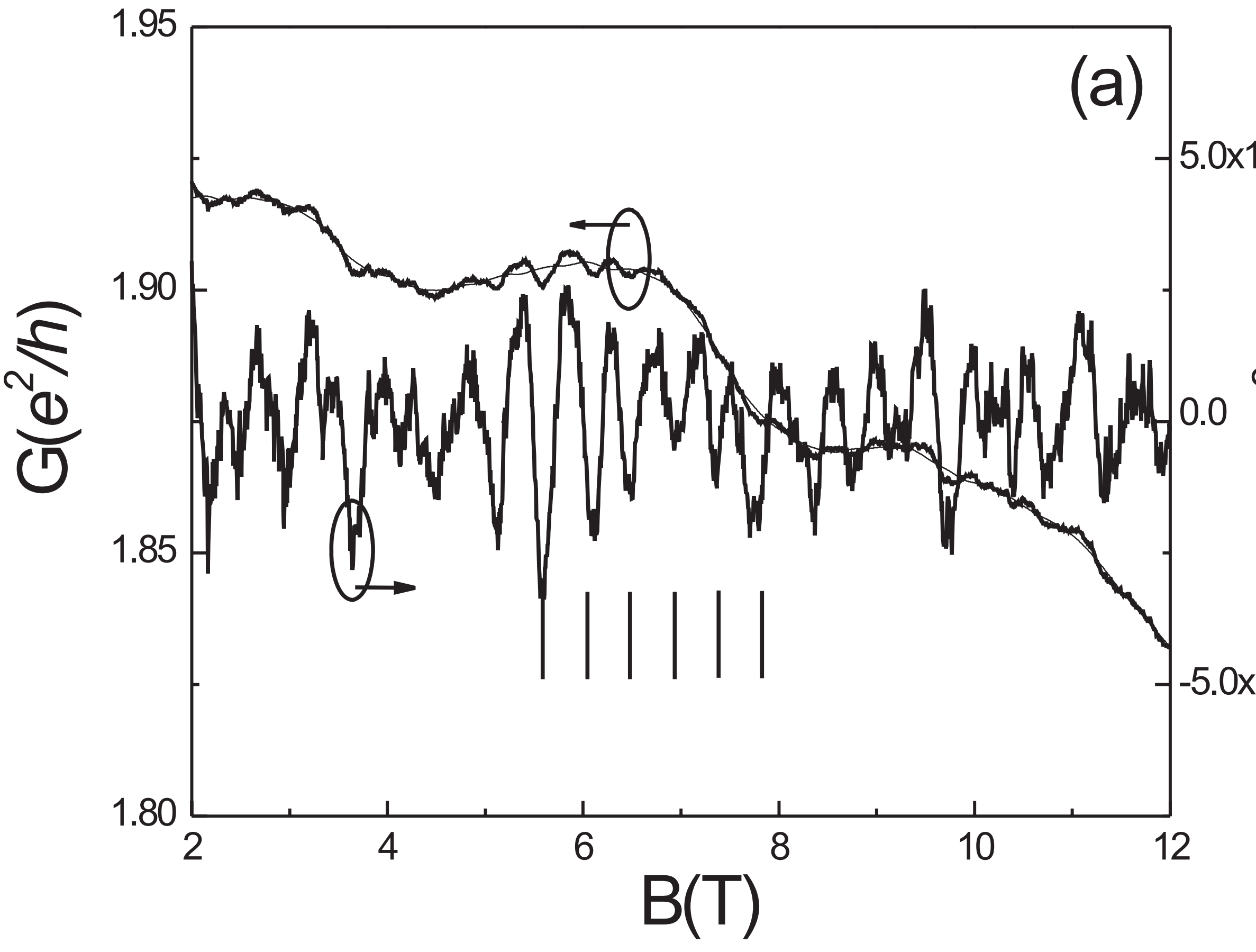

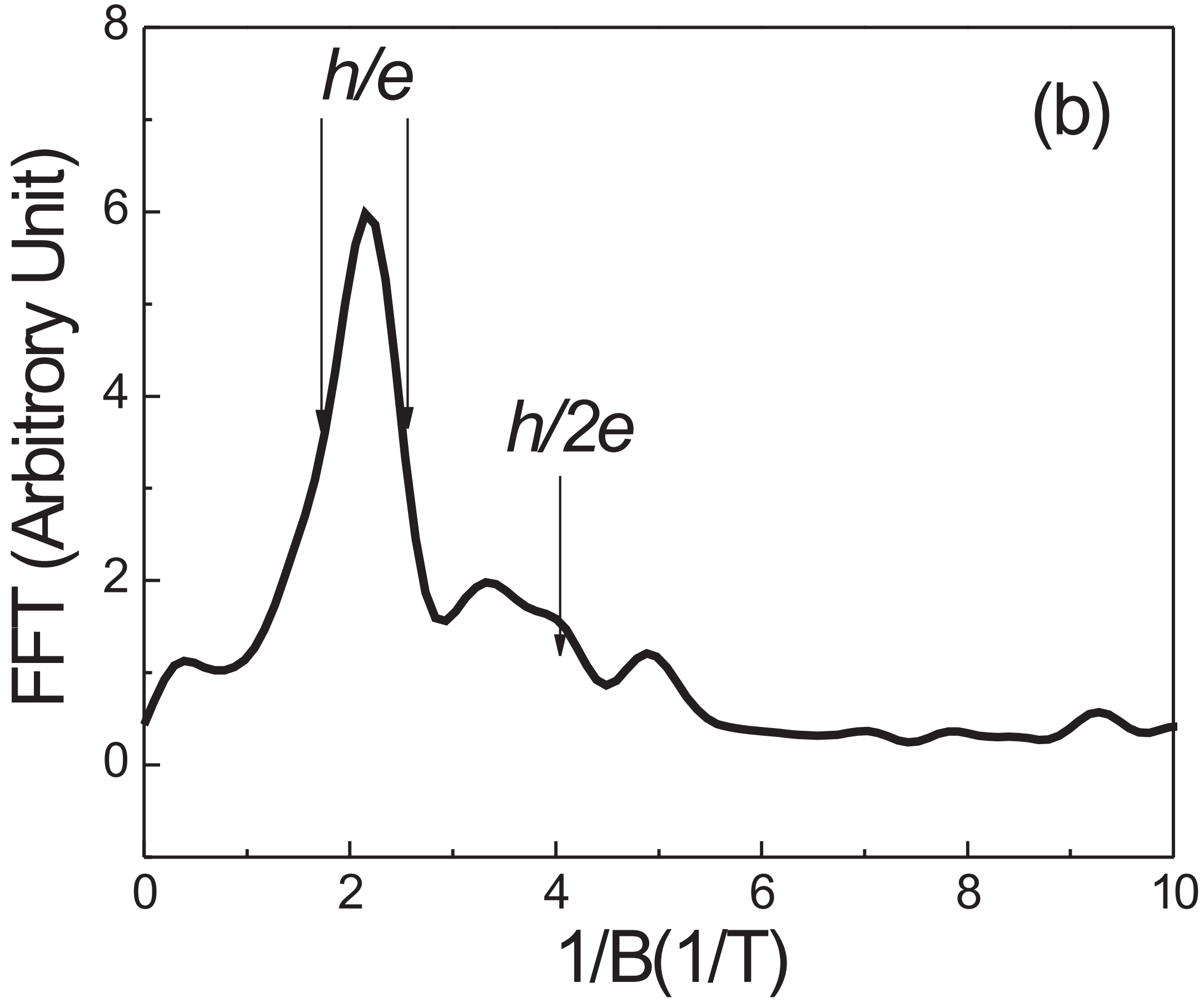

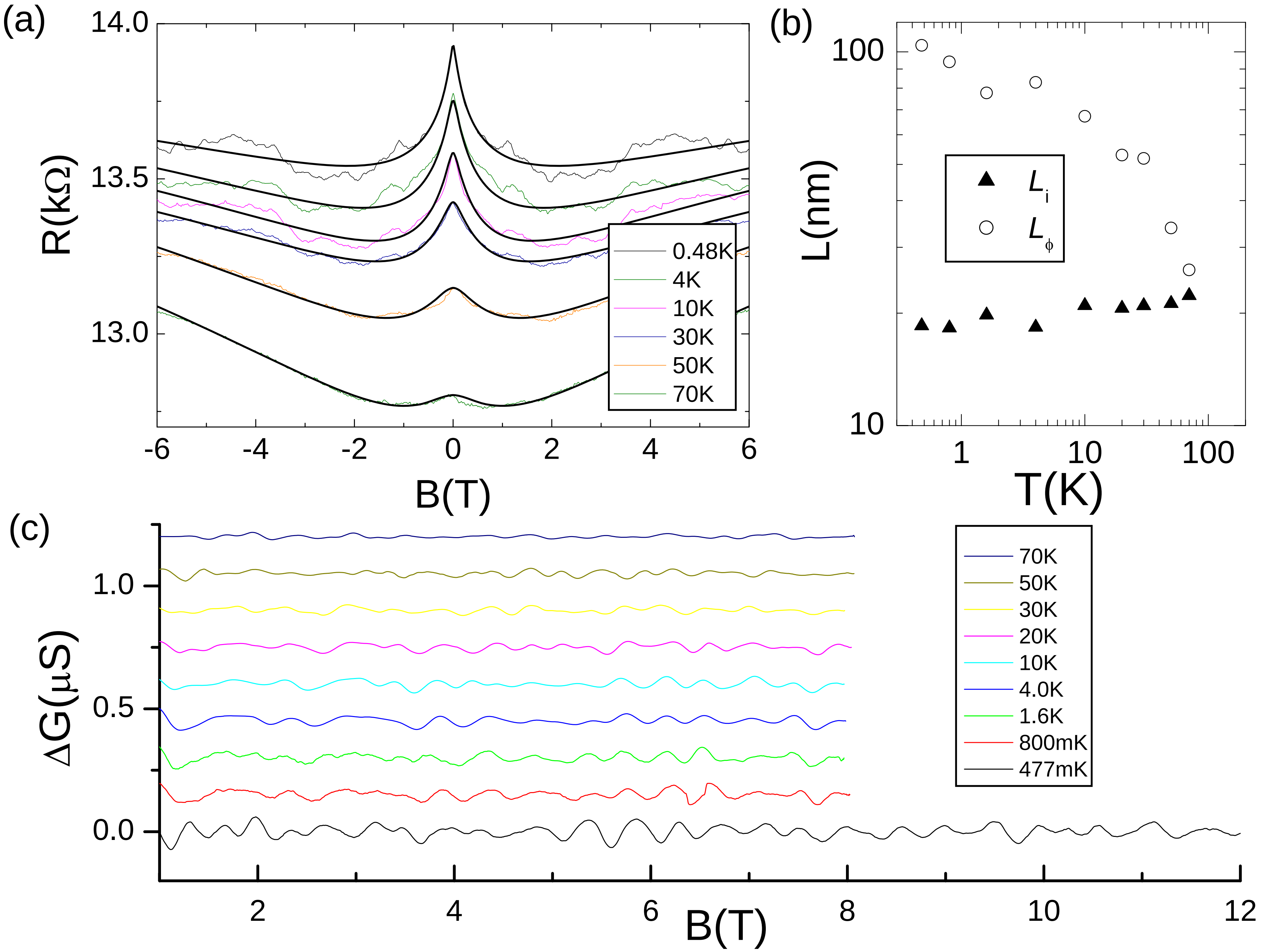